\documentclass[prl,twocolumn,showpacs,floatfix,superscriptaddress,citeautoscript]{revtex4}

\usepackage{graphicx}
\usepackage{dcolumn}   
\usepackage{bm}        
\usepackage{amssymb}   
\usepackage{verbatim}
\usepackage{multirow}

\usepackage{amsmath}
\usepackage{amsfonts}

\begin{document}

\title{Microwave response of vortices in superconducting thin films of Re and Al}

\author{C. Song}
\affiliation{Department of Physics, Syracuse University, Syracuse, NY 13244-1130}
\author{T.W. Heitmann}
\affiliation{Department of Physics, Syracuse University, Syracuse, NY 13244-1130}
\author{M.P. DeFeo}
\affiliation{Department of Physics, Syracuse University, Syracuse, NY 13244-1130}
\author{K. Yu}
\affiliation{Department of Physics, Syracuse University, Syracuse, NY 13244-1130}
\author{R. McDermott}
\affiliation{Department of Physics, University of Wisconsin-Madison, Madison, WI 53706}
\author{M. Neeley}
\affiliation{Department of Physics, University of California, Santa Barbara, CA 93106}
\author{John M. Martinis}
\affiliation{Department of Physics, University of California, Santa Barbara, CA 93106}
\author{B.L.T. Plourde}
\email[]{bplourde@phy.syr.edu}
\affiliation{Department of Physics, Syracuse University, Syracuse, NY 13244-1130}

\date{\today}


\begin{abstract}
Vortices in superconductors driven at microwave frequencies exhibit a response related to the interplay between the vortex viscosity, pinning strength, and flux creep effects. At the same time, the trapping of vortices in superconducting microwave resonant circuits contributes excess loss and can result in substantial reductions in the quality factor. Thus, understanding the microwave vortex response in superconducting thin films is important for the design of such circuits, including superconducting qubits and photon detectors, which are typically operated in small, but non-zero, magnetic fields. By cooling in fields of the order of $100$~$\mu$T and below, we have characterized the magnetic field and frequency dependence of the microwave response of a small density of vortices in resonators fabricated from thin films of Re and Al, which are common materials used in superconducting microwave circuits. Above a certain threshold cooling field, which is different for the Re and Al films, vortices become trapped in the resonators. Vortices in the Al resonators contribute greater loss and are influenced more strongly by flux creep effects than in the Re resonators. This different behavior can be described in the framework of a general vortex dynamics model. 
\end{abstract}

\maketitle

\section{Introduction}

Superconducting thin films in sufficiently large magnetic fields are threaded by vortices of quantized magnetic flux that interact with currents flowing in the films as well as materials defects.  
The microwave response of superconductors can be profoundly influenced by the presence of vortices and the dynamics they exhibit at high frequencies. 
This can play an important role in the design of superconducting microwave devices, where 
the ability to fabricate low-loss resonant circuits has enabled the development of entirely new classes of experiments.
There have been many recent groundbreaking studies of quantum coherent superconducting circuits, which could serve as qubits for forming the elements of a quantum computer \cite{clarke08}. In addition, there has been much progress in the development of superconducting Microwave Kinetic Inductance Detectors (MKIDs), which are highly sensitive photon detectors for astrophysical measurement applications \cite{zmuidzinas03}. 
In a quest to improve the performance of such circuits yet further, there have been many recent efforts to probe the effects of microwave loss in a variety of areas, including dielectric layers and film surfaces that form the microwave superconducting circuits \cite{martinis05, gao08, barends08}.
Another possible loss mechanism in these systems is the dissipation due to vortices trapped in the superconducting traces, which
can result in substantial reductions in the quality factor of superconducting resonators.  
Thus, understanding this dissipation mechanism is important for the design of microwave superconducting circuits.

Because superconductivity is suppressed in the core of a vortex, the motion of vortices leads to dissipation, as is typically characterized by the flux-flow vortex viscosity $\eta = \Phi_0 B_{c2}/\rho_n$, where $\Phi_0$ is the flux quantum, $h/2e$, $\rho_n$ is the normal-state resistivity of the material, and $B_{c2}$ is the upper-critical field \cite{tinkham96, stephen65}. 
A current density ${\bf J}$ flowing through a superconductor exerts a Lorentz force on the vortices, ${\bf F}_L = {\bf J} \times \Phi_0 \hat{n}$.
Thus, an oscillatory current in a superconducting microwave circuit can generate dissipative vortex motion.
However, any practical superconductor inherently contains various materials defects which produce vortex pinning, thus complicating the situation. 
In the simplest case, the pinning potential wells $U(x)$ can be assumed to be harmonic with spring constant $k_p$, giving a pinning force ${\bf F}_p = k_p {\bf x}$. The vortex equation of motion at zero temperature is given by
\begin{equation}
\eta \dot{\bf x} + k_p {\bf x} = {\bf F}_L,
\label{eq:vortex-eom}
\end{equation}
where we have neglected a possible vortex mass, which likely would not play a role until much higher frequencies than the circuits in our experiments that operate between $\sim 2-11$~GHz  \cite{Suhl65}.
Thus, the interplay between the viscous force and the pinning will determine the frequency dependence of the vortex response. At low frequencies the pinning will dominate and the response will be primarily elastic, while at higher frequencies, the viscosity will become more important and the response will be more dissipative.

In one of the original investigations of the microwave dynamics of vortices in superconductors, Gittleman and Rosenblum measured circuits patterned from PbIn and NbTa foils and described the vortex response in terms of Eq. (\ref{eq:vortex-eom}) \cite{gittleman66, gittleman68}. 
Similar measurements were also performed on Al thin films in Ref. \cite{possin68}. 
Several decades later, various groups studied vortices at microwave frequencies in YBCO films, with particular relevance to high-T${\rm _c}$ thin-film microwave devices  \cite{belk96, revenaz94, powell98, pompeo08}.
Recently there have also been investigations of the microwave vortex dynamics in MgB$_2$ \cite{zaitsev07, ghigo07, sarti05} and Nb films \cite{janjusevic06}.
Previous work on the microwave response of vortices in superconductors has primarily involved large magnetic fields, at least several orders of magnitude larger than the Earth's field. On the other hand, superconducting resonant circuits for qubits and detectors are typically operated in relatively small magnetic fields, of the order of 
$100$~$\mu$T or less and are fabricated from low-T$_c$ thin films that are often type-I superconductors in the bulk. 
In this article, we report on measurements probing the magnetic field and frequency dependence of the microwave response of a small number of vortices using resonators fabricated from thin films of rhenium and aluminum -- common materials used in superconducting resonant circuits for qubits and detectors. Related measurements that motivated the present work were performed in Ref. \cite{mcdaniel06}.

\begin{figure}
\centering
  \includegraphics[width=3.35in]{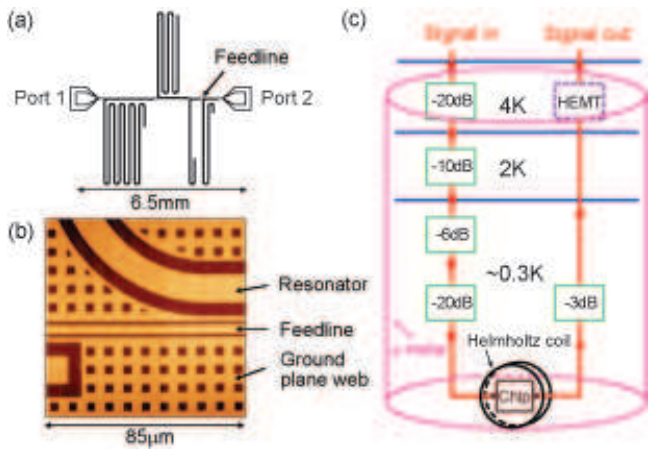}
  \caption{(Color online) (a) Chip layout showing common feedline and four resonators. (b) Atomic Force Microscope (AFM) image of portion of Al chip. (c) Schematic of measurement setup, including cold attenuators with values listed in dB.
\label{fig:setup}}
\end{figure}

\section{Resonator design and measurement procedure}

The use of resonant circuits allows us to probe small changes in the response due to the introduction of a few vortices. Several resonant circuit geometries are possible, but the 
coplanar waveguide (CPW) geometry is particularly straightforward for implementing with thin films and is a common configuration used in superconducting qubit and MKID circuits.
In order to map out the frequency dependence of the vortex response, we would like to have multiple resonators of different lengths patterned from the same film that we measure under the same magnetic field and temperature conditions. 
Such an arrangement is possible with a similar multiplexing scheme to what was developed recently for MKIDs, with multiple quarter-wave resonators of different lengths capacitively coupled to a common feedline \cite{mazin04,zmuidzinas03}. 

Our layout consists of four quarter-wave CPW resonators with lengths $15.2$, $9.3$, $4.4$, and $2.8$~mm, which, if we neglect the effects of the kinetic inductance of the superconductors for now, yields fundamental resonances near $1.8$, $3.3$, $6.9$, and $11.0$ GHz, as calculated with the Sonnet microwave circuit simulation software \cite{sonnet05}.
As with the resonators, our feedline also has a CPW layout, with a nominal impedance of $50$~$\Omega$, and runs across the centerline of the chip.
Each resonator follows a serpentine path in order to fit on the chip, with an elbow bend at the open end, while the opposite end is shorted to the ground plane.
The coupling capacitance between each resonator and the feedline is determined by the length of its elbow. 
We design for the resonators to be somewhat over-coupled at zero field, where the loss at the measurement temperatures is dominated by thermally-excited quasiparticles. This gives us the ability to continue to resolve the resonance lines with the anticipated enhanced levels of loss once vortices are introduced. 

In order to control the number of vortices in the resonators, we cool through the transition temperature $T_c$ in an applied magnetic field $B$. 
The process for the trapping of vortices in a thin superconducting strip of width $w$ upon field-cooling has been studied experimentally \cite{stan04,kuit08,bronson06} and theoretically  \cite{clem98,maksimova98,likharev72}, indicating
a threshold cooling field $B_{th}$ below which all of the magnetic flux will be expelled from the strip. Apart from numerical details of the various approaches, 
this threshold field has been shown to scale approximately like $B_{th} \sim \Phi_0/w^2$. 
In order to trap vortices only in the resonators, we design the ground plane to have a lattice of holes, with the webbing and the feedline linewidth to be a factor of three narrower than that of the resonator, which is nominally $12$~$\mu$m [Fig. \ref{fig:setup}(b)].
This should then provide about a decade of range in the cooling field where vortices are primarily trapped in the resonators, with
$\Phi_0/w^2 \approx 14$~$\mu$T. 

We use the same layout from Fig. \ref{fig:setup}(a) to pattern resonators from thin films of Re and Al. The Re films were $50$~nm thick and were deposited by electron-beam evaporation onto a-plane sapphire at a temperature of $850$~C. The Al films were $150$~nm thick and were also electron-beam evaporated onto c-plane sapphire that was not heated. Both types of films were patterned photolithographically followed by a reactive ion etch in a combination of BCl$_3$, Cl$_2$, and CH$_4$ (Al) or SF$_6$ and Ar (Re). The superconducting transition temperatures $T_c$ for the films were identified with the corresponding step in the microwave transmission $S_{21}$ through the feedline away from any of the resonance dips, leading to $T_c^{\rm Re}=1.70$~K and $T_c^{\rm Al}=1.13$~K.
The width of the center conductors for the measured resonators was $11.9~\mu$m for the Re and $11.5~\mu$m for the Al.  
The normal state resistivities were measured to be $\rho_n^{\rm Re}=1.6$~$\mu \Omega$-cm at $4$~K, with RRR~$=11$, for the Re, and $\rho_n^{\rm Al}=0.33$~$\mu \Omega$-cm at $2$~K, with RRR~$=10$, for the Al.
In the bulk, both Re and Al are Type-I superconductors, 
however, films of Type-I superconductors with thicknesses less than the bulk coherence length in perpendicular magnetic fields have been shown to support the nucleation of $h/2e$ Abrikosov vortices \cite{tinkham63,huebener01,fetter67,maloney72,dolan74}.

We cool the resonators to $\sim 300$~mK using a ${\rm ^3}$He refrigerator and we generate the magnetic field with a superconducting Helmholtz coil. A $\mu$-metal cylinder attenuates stray magnetic fields in the laboratory. 
We perform our measurements using a vector network analyzer (Agilent N5230A) to record the magnitude and phase of the transmission through our feedline, $S_{21}$. 
The microwave drive signal is delivered to one side of the feedline through a lossy stainless steel semi-rigid coaxial line combined with -56 dB of cold attenuation [Fig. \ref{fig:setup}(c)]. 
The chip containing the resonators is mounted and wirebonded into a custom chip carrier with ports for transmitting signals through the feedline.
The signal on the output side of the feedline is amplified with a cryogenic HEMT amplifier (Caltech, model 165D) that is mounted on the 4K flange of the refrigerator, before returning to the network analyzer.
The HEMT has a gain of $\sim 38$~dB between $0.5 - 11$~GHz with a noise temperature in this range of $T_N \approx 5$~K.
We also include a $3$~dB attenuator between the output of the chip carrier and the coaxial line to the HEMT to suppress spurious resonances and reduce noise fed back from the HEMT input; a circulator would not be practical here because of the wide bandwidth required for our measurements.

\begin{figure}
\centering
  \includegraphics[width=3.35in]{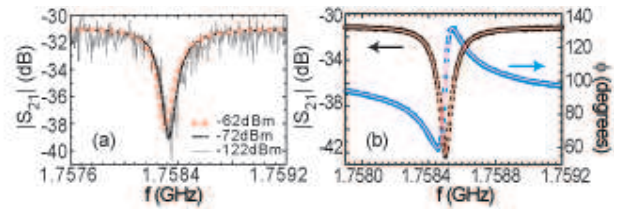}
  \caption{(Color online) (a) Dips in magnitude of $S_{21}$ for different microwave drive power for the Re resonator near $1.8$~GHz, $B=92.5$~$\mu$T. (b) Magnitude and phase 
  of $S_{21}$ for Re resonator near $1.8$~GHz cooled in $B=56.4$~$\mu$T (symbols), along with the corresponding fit (solid lines) as described in text; $Q_{fit} = 10040$, $f_0 = 1.758499$~GHz.
  \label{fig:power&fits}}
\end{figure}

In the vicinity of a resonance we observe a dip in $|S_{21}|$, while away from the resonances, the feedline exhibits full transmission, upon accounting for the cold attenuation, loss from the chip carrier and stainless steel coaxial line, as well as the gain from the HEMT. The response of these various components was calibrated separately with our network analyzer. 
Over a wide range of power, roughly $60$ dB, we observe no variation of the resonance lineshape [Fig. \ref{fig:power&fits}(a)]. For stronger driving, $\sim -62$~dBm  or larger delivered to the feedline, the dip becomes nonlinear and the quality factor decreases. The nonlinear response of strongly driven superconducting resonators has been investigated extensively in a variety of contexts \cite{abdo07,golosovsky95,chin92}.
To avoid such strong-driving nonlinearities we measure our resonators with a weak microwave drive, typically delivering a power of less than $-82$~dBm to the feedline.
In order to extract the quality factor $Q_{fit}$ and center frequency $f_0$ for each resonator, we fit the resonance trajectory in the complex plane, following a similar $10$-parameter fitting procedure to what is done for MKID measurements \cite{mazin04}. 
Figure \ref{fig:power&fits}(b) shows an example measurement of the magnitude and phase of $S_{21}$ for the Re resonator near $1.8$~GHz cooled in $B=56.4$~$\mu$T, along with the corresponding fit. 

\section{Measurements}

We study the influence of vortices in the resonators by repeatedly field-cooling through ${\rm T_c}$ in different magnetic fields. 
For each value of $B$, we heat the sample above $T_c $ to $1.95$~K ($1.4$~K) for Re (Al), adjust the current through our Helmholtz coil to the desired value, then cool down to $300$~mK ($310$~mK) for Re (Al). The cooling time for each field point is approximately $30$ minutes. 
During our measurements we regulate the temperature on the sample stage to within $\pm 0.2$~mK of the stated values.

\begin{figure}
\centering
  \includegraphics[width=3.35in]{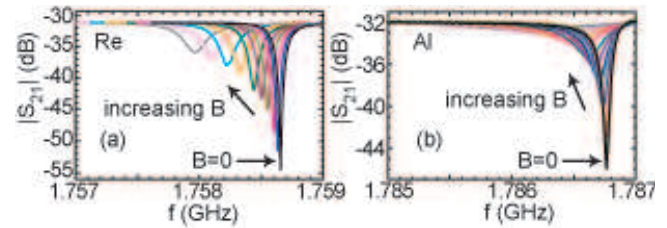}
  \caption{(Color online) Magnitude of $S_{21}$ for different cooling fields $B$ for resonator near $1.8$~GHz on (a) Re chip with $B$ from $0$ to $149.6$~$\mu$T; (b) Al chip with $B$ from $0$ to $94.5$~$\mu$T.
 \label{fig:re&al-dips}}
\end{figure}

The addition of vortices through field-cooling results in a downward shift in the resonance frequency and a reduction in the quality factor.
This general trend can be seen in Figure \ref{fig:re&al-dips} where we plot the magnitude of $S_{21}$ for several different cooling fields for the Re and Al chips for the resonator near $1.8$~GHz. 
While the general trend is similar for the Re and Al resonators, the details of the response for the two materials are clearly quite different, with a more substantial broadening of the resonance dip with $B$ for Al compared to the Re.
By fitting the resonance trajectories for each of the four resonators at each cooling field on the Re and Al chips, we are able to extract the field and frequency dependence of $Q_{fit}$ and $f_0$ for the two materials.

\begin{figure}
\centering
 \includegraphics[width=3.35in]{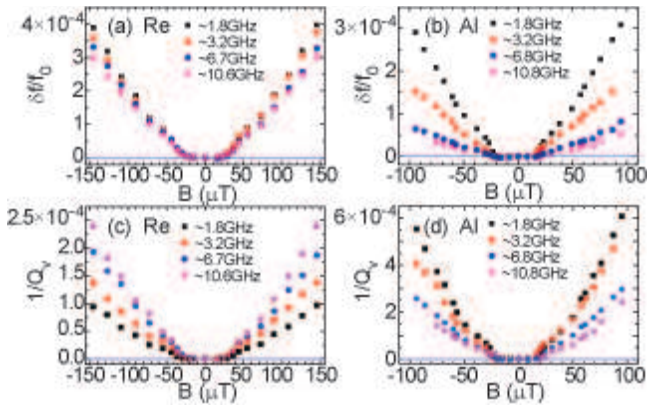}
\caption{(Color online) Fractional frequency shift $\delta f/f_0 (B)$ for (a) Re and (b) Al. Excess loss due to vortices $1/Q_v (B)$, as defined by Eq. (\ref{eq:excess-loss}) for (c) Re and (d) Al.
\label{fig:loss&shift-all}}
\end{figure}

We compute the excess loss in each resonator due to the presence of vortices, $1/Q_v$, by fitting the resonance at a particular magnetic field to obtain $1/Q_{fit}(B)$ and subtracting the inverse quality factor measured with $B = 0$, according to
\begin{equation}
1/Q_v = 1/Q_{fit}(B) - 1/Q_{fit}(0),
\label{eq:excess-loss}
\end{equation}
thus removing the loss due to thermal quasiparticles, coupling to the feedline, and any other field-independent loss mechanisms. 
The uncertainties in our values of $1/Q_v$ from the fitting process are less than $7 \times 10^{-6}$ and the corresponding error bars are too small to be seen for most of the points in Figure \ref{fig:loss&shift-all}(c, d).
 
We extract the fractional frequency shift of each resonance relative to its center frequency at $B = 0$:
\begin{equation}
\delta f/f_0 =\left[f_0(0)-f_0(B)\right]/f_0(0).
\label{eq:freq-shift}
\end{equation}
The uncertainties in our values of $\delta f/f_0$ from the fitting process are less than $2 \times 10^{-10}$ and are not visible in Figures \ref{fig:loss&shift-all}(a, b).

We plot $1/Q_v (B)$ and $\delta f/f_0 (B)$ for Re [Figs. \ref{fig:loss&shift-all}(a, c)] and Al [Figs. \ref{fig:loss&shift-all}(b, d)].
For both materials, there is a region near zero field where there is essentially no change in $1/Q_v$ or $f_0$, corresponding to cooling fields below the threshold for trapping vortices in the resonators. 
Above this threshold, both $1/Q_v$ and $\delta f/f_0$ increase with $|B|$ for both materials. However, the frequency dependences of these quantities are quite different between the Re and Al films.
For the Re resonators at a particular $B$, $\delta f/f_0$ decreases slightly with increasing frequency [Fig. \ref{fig:loss&shift-all}(a)], while for Al there is a substantial decrease in $\delta f/f_0$ with increasing frequency [Fig. \ref{fig:loss&shift-all}(b)].
Even more striking, the loss due to vortices $1/Q_v$ increases with frequency for Re [Fig. \ref{fig:loss&shift-all}(c)], while it decreases for Al [Fig. \ref{fig:loss&shift-all}(d)].

\section{Analysis and Effective Resistivity}

Gittleman and Rosenblum (GR) first considered Eq. (\ref{eq:vortex-eom}) and derived a complex resistivity due to the vortex response \cite{gittleman68}. 
This model was later extended by Coffey and Clem \cite{coffey91}, as well as Brandt \cite{brandt91}, 
to address issues of microwave vortex dynamics in the high-T$_c$ superconductors, including the influence of flux creep, where vortices can wander between pinning sites either by thermal activation or tunneling \cite{blatter94}. 
Pompeo and Silva demonstrated that these various models can be described by a single expression for an effective complex resistivity $\tilde{\rho}_v$ due to vortices \cite{pompeo08}:
\begin{equation}
\tilde{\rho}_v = \frac{\Phi_0 \left( B-B_{th} \right)}{\eta_{e}} \frac{\epsilon + i f/f_d}{1+ i f/f_d},
\label{eq:pompeo-silva}
\end{equation}
where $f_d = k_p/2\pi \eta_e$ is the characteristic depinning frequency that corresponds to the crossover from elastic to viscous response; $\epsilon$ is a dimensionless quantity that describes the strength of the flux creep and can range between $0$ -- recovering the zero-temperature GR model -- and $1$ -- when $\tilde{\rho}_v$ is purely real and equal to the conventional Bardeen-Stephen flux-flow resistivity \cite{bardeen65}. 
The threshold cooling field is accounted for by including $B_{th}$. For $B<B_{th}$ there are no vortices present and $\tilde{\rho}_v = 0$, although pinning can result in the trapping of vortices for $B$ somewhat smaller than $B_{th}$.
The real part of $\tilde{\rho}_v$ is associated with the loss contributed by the vortices, while the imaginary part of $\tilde{\rho}_v$ determines the reactive response of the vortices. 
Relating $f_d$ and $\epsilon$ to the 
pinning potential depends on the details of the particular vortex dynamics model one considers \cite{pompeo08}.

In a variety of contexts the microwave response of a superconductor is often characterized in terms of the surface impedance $Z_s = R_s+i X_s$. 
Changes in $Z_s$ under different conditions, for example, different vortex densities determined by $B$, can then be separated into changes in the surface resistance $\Delta R_S(B)$ and reactance $\Delta X_S(B)$, where these quantities correspond to the differences between measurements at $B$ and zero field. 
For a particular superconducting resonator, $\Delta R_S(B)$ and $\Delta X_S(B)$ can be related to the observable quantities $1/Q_{fit}$ and $\delta f/f_0$ through
\begin{eqnarray}
\Delta R_s(B) &=& G \Delta \left[ 1/Q_{fit}(B) \right] = G \left[ 1/Q_v(B)\right], \\
\Delta X_s(B) &=& 2 G \left[ \delta f/f_0(B) \right],
\label{eq:surf-imped}
\end{eqnarray}
where the geometrical parameter $G$ depends on the details of the resonator geometry, the current distribution, and the kinetic inductance contribution \cite{zaitsev03, hein99}. 
Often the dimensionless ratio $r = \Delta X_S/\Delta R_S$ lends useful insight into the microwave response and 
thus eliminates the influence of $G$.
The complex vortex resistivity $\tilde{\rho}_v$ can also be related to $r$ as
\begin{equation}
r = \frac{{\rm Im}(\tilde{\rho}_v)}{{\rm Re}(\tilde{\rho}_v)},
\label{eq:r-param}
\end{equation}
thus providing a path for comparing our measured quantities with the generalized vortex response given by Eq. (\ref{eq:pompeo-silva}) \cite{zaitsev07, pompeo08}. By analyzing the $r-$parameter and its frequency dependence from our measurements, we will extract $f_d$ and $\epsilon$ for the Re and Al films. We can then study the field dependence of the loss or frequency shift data separately to compare $\eta_e$ for the two materials.

\begin{figure}
\centering
 \includegraphics[width=3.35in]{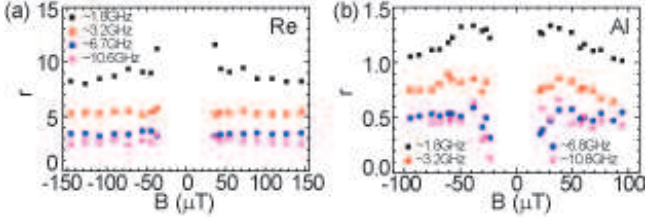}
\caption{(Color online) $r(B)$ for (a) Re and (b) Al films for four different resonator lengths.
\label{fig:r-param-field}}
\end{figure}

In Figure \ref{fig:r-param-field} we plot $r(B)$ calculated from the data in Figure \ref{fig:loss&shift-all} for the four different resonators on the Re and Al chips. 
For the Re resonators $r$ is well above unity, indicating the dominance of the reactive contribution of the vortex dynamics in the frequency range covered by our chip layout. In contrast, $r$ is near or somewhat less than unity for the Al resonators, indicating the significant loss related to the vortex motion in this system. 
When $|B|$ is less than the threshold to trap vortices, $1/Q_v \approx 0$ and $r$ diverges, thus we do not include values for $r$ in this range in Fig. \ref{fig:r-param-field}.  
For $|B|$ somewhat larger than the threshold, $r$ becomes roughly field-independent, particularly for the Re film. 
When $|B|$ is just beyond the threshold, there are clear differences in $r(B)$ between the Re and Al films that will be addressed shortly.

\begin{figure}
\centering
\includegraphics[width=3.35in]{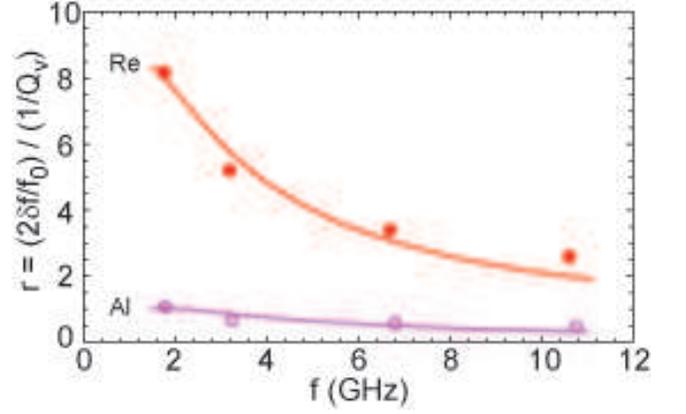}
\caption{(Color online) $r(f)$ for Re with $B= -130.9$~$\mu {\rm T}$ (closed circles) and Al with $B= -75.5$~$\mu {\rm T}$ (open circles) along with fits as described in text. Fit parameters are $f_d=22.6$~GHz ($4.2$~GHz) and $\epsilon=0.0039$ ($0.15$) for Re (Al).
\label{fig:r-param-fits}}
\end{figure}

\begin{figure}
\centering
\includegraphics[width=3.35in]{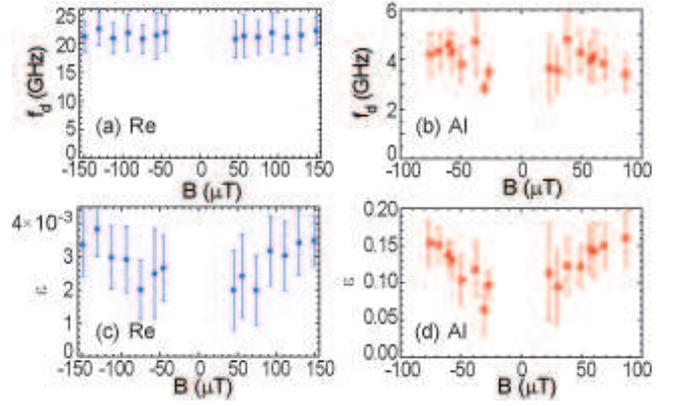}
\caption{(Color online) $B-$dependence of parameters from fits to $r(f)$ at each $B$: $f_d$ for (a) Re and (b) Al; $\epsilon$ for (c) Re and (d) Al films. Note the different scale factors on $\epsilon$ between (c) and (d). 
\label{fig:depin&creep}}
\end{figure}

The frequency dependence of $r$ can be seen in Figure \ref{fig:r-param-field} by focusing on a particular value of $B$ and observing the variation in $r$ for the four resonators. We plot this explicitly in Figure \ref{fig:r-param-fits} for one field each for Re and Al, where, for both films, $r$ decreases with frequency. 
We can make a two-parameter fit to the $r(f)$ data in Fig. \ref{fig:r-param-fits} with Eqs. (\ref{eq:pompeo-silva}, \ref{eq:r-param}) by varying $f_d$ and $\epsilon$. 
Performing this same analysis for each value of $B$ in Figure \ref{fig:r-param-field} yields fit values $f_d(B)$ and $\epsilon(B)$ (Fig. \ref{fig:depin&creep}). We note that for both our Re and Al data, it is not possible to fit $r(f)$ with $\epsilon=0$.

From Figure \ref{fig:depin&creep}, there is clearly a substantial difference in $f_d$ for the Re and Al films. 
For $|B|>50$~$\mu$T, well beyond the threshold for trapping vortices, the average of $f_d^{\rm Re}$ from Figure \ref{fig:depin&creep}(a) is $22$~GHz, much higher than our highest resonator fundamental frequency. 
In contrast, for Al, the average of $f_d^{\rm Al}$ from Figure \ref{fig:depin&creep}(b) is $4$~GHz, near the lower end of our resonator frequencies. 

The ratio of the depinning frequencies $f_d^{\rm Re}/f_d^{\rm Al}$ can be used to compare the relative pinning strength for the Re and Al films with the following expression
\begin{equation}
\frac{k_p^{\rm Re}}{k_p^{\rm Al}} = \left( \frac{f_d^{\rm Re}}{f_d^{\rm Al}} \right) \left( \frac{\eta_e^{\rm Re}}{\eta_e^{\rm Al}} \right).
\label{eq:fd-ratio}
\end{equation}

We can extract $\eta_e^{\rm Re}$/$\eta_e^{\rm Al}$ from the $1/Q_v(B)$ data of Figure \ref{fig:loss&shift-all} based on Eq. (\ref{eq:pompeo-silva}) by writing the resistance due to vortices $R_v$ as 
\begin{equation}
R_v = j(x) {\rm Re} \left[ \tilde{\rho}_v \right] \left( l/w t\right),
\label{eq:vortex-resistance}
\end{equation}
where $l$ is the resonator length, $t$ is the thickness, and $j(x)$ is a dimensionless factor that scales $R_v$ based on the current density $J_s(x)$ at the position $x$ of the vortices across the width of the resonator. In general, $J_s(x)$ will be non-uniform with more current flowing along the edges of the center conductor. Thus, $j(x)=J_s(x)^2/\langle J_s \rangle^2$, where $\langle J_s \rangle$ is the average current density across the center conductor. The numerical calculation of $j(x)$ will be discussed further in the subsequent section. 
For a resonator at $f_0$, $1/Q_v$ can be related to $R_v$ and then $\tilde{\rho}_v$ by
\begin{eqnarray}
1/Q_v  &=&  \left( R_v/l \right)/2 \pi f_0 L' \\
 &=&  j(x) {\rm Re} \left[ \tilde{\rho}_v \right]/2 \pi f_0 w t L' \label{eq:f-over-Q}
\end{eqnarray}
where $L'$ is the inductance per unit length of the resonator. 
After applying the definition of $\tilde{\rho}_v$ from Eq. (\ref{eq:pompeo-silva}) we then differentiate both sides of Eq. (\ref{eq:f-over-Q}) with respect to $B$:
\begin{equation}
\frac{\partial (1/Q_v)}{\partial B}  = \frac{ j(x) \Phi_0}{2 \pi f_0 w L'} \left( \frac{1}{t  \eta_e} \right) \left[ \frac{\epsilon+(f_0/f_d)^2}{1+(f_0/f_d)^2} \right].
\label{eq:f-over-loss-scaled}
\end{equation}

\begin{figure}
\centering
\includegraphics[width=3.35in]{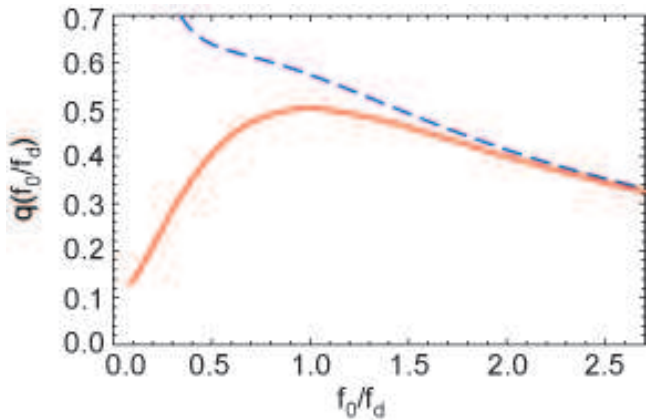}
\caption{(Color online) Plot of $q(f_0/f_d)$, computed based on definition in the text, which is proportional to $\partial (1/Q_v)/\partial B$ with $\epsilon$ values from text corresponding to Re (solid red line) and Al (dashed purple line).
\label{fig:loss-vs-freq}}
\end{figure}

By scaling with the frequency-independent factors on the right-hand side of Eq. (\ref{eq:f-over-loss-scaled}), we can investigate the frequency dependence of $\partial (1/Q_v)/\partial B$. In Figure \ref{fig:loss-vs-freq} we plot $q(f_0/f_d) = (f_d/f_0) (\epsilon+(f_0/f_d)^2)/(1+(f_0/f_d)^2)$ for the $\epsilon$ values obtained from fits to the $r(f)$ data for Re and Al. With $\epsilon$ small, $q(f_0/f_d)$ is an increasing function for $f_0 < f_d$ -- characteristic of our measurements on Re, where all of the resonances are below $f_d^{\rm Re}$ and there is greater loss at higher frequencies. For $f_0 > f_d$, $q(f_0/f_d)$ is a decreasing function. In addition, a larger value of $\epsilon$ enhances the loss at frequencies comparable to and less than $f_d$, and this dependence is characteristic of our measurements on Al, where $f_d^{\rm Al}$ is near the lower end of our resonances and we observe a decrease in the loss for increasing frequency.

We can compute $\partial \left( 1/Q_v\right)/\partial B$ from the data in Fig. \ref{fig:loss&shift-all}(c, d) by making linear fits for the data beyond the threshold shoulder. If we then use the $\epsilon$ and $f_d$ parameters from the $r(f)$ fits in Figure \ref{fig:depin&creep}, and neglect the small difference in $L'$ between the Re and Al resonators because of differences in kinetic inductance, 
we can apply Eq. (\ref{eq:f-over-loss-scaled}) to the Al and Re data, then take the ratio of these for each of the four resonator lengths. After accounting for $t^{\rm Re}/t^{\rm Al}$ we obtain 
$\eta_e^{\rm Re}/\eta_e^{\rm Al} \approx 1$. A related analysis involving the $\delta f/f_0$ data and ${\rm Im}[\tilde{\rho}_v]$ yields approximately the same value for $\eta_e^{\rm Re}/\eta_e^{\rm Al}$.

For comparison, we can also estimate $\eta_e^{\rm Re}$/$\eta_e^{\rm Al}$  assuming $\eta_e$ corresponds to the
Bardeen-Stephen (BS) flux-flow viscosity $\eta$ \cite{bardeen65}. In this model, each vortex core is treated as a normal cylinder with a radius equal to the effective coherence length $\xi_e$ with resistivity $\rho_n$. Dissipation during the vortex motion leads to a viscosity $\eta = \Phi_0^2/2 \pi \rho_n \xi_e^2$ \cite{bardeen65, tinkham96}. 

Using $\rho^{\rm Al} l^{\rm Al} = 4 \times 10^{-16}$~$\Omega$~m$^2$ from Ref. \cite{romijn82} and our measured value of $\rho^{\rm Al}$, we estimate the electronic mean free path of our Al film to be of the order of $100$~nm, 
much less than the BCS coherence length for Al, $\xi_0 \approx 1500$~nm \cite{tinkham96}, thus putting the Al film well into the dirty-limit. We have measured the shift in $T_c$ as a function of magnetic field for the Al film, and thus obtained $S=-dB_{c2}/dT\mid_{T_c}$, which we can then use with the standard dirty-limit expression \cite{kes83} to obtain the effective coherence length $\xi_e^{\rm Al} \approx 230$~nm, consistent with estimates for other Al thin films \cite{romijn82}.

We are not aware of any measurements of the coherence length in Re. Furthermore, it is not clear if the Re films are in the dirty limit, thus we can attempt to estimate $\xi_e^{\rm Re}$ using the BCS expression: $\xi_0 = \hbar v_F/\pi \Delta(0)$ with $\Delta(0)=1.76 k_B T_c$ \cite{tinkham96}. If we apply the free-electron model, we can write $v_F=(\pi k_B/e)^2/\gamma \rho l$, where $\gamma$ is the linear coefficient of the specific heat ($260$~J m$^{-3}$~K$^{-2}$ for Re \cite{mcmillan68}). The quantity $\rho l$ is the product of the resistivity and mean free path with reported values for Re of $4.5 \times 10^{-15}$~$\Omega$~m$^2$ in Ref. \cite{ulhaq82} and $2.16 \times 10^{-15}$~$\Omega$~m$^2$ in Ref. \cite{tulina82}. This results in $\xi_0^{\rm Re} \approx 50-100$~nm. We note that these values of $\rho l$ imply a mean free path for our Re film between $\sim 140-280$~nm, thus confirming that the film is not in the dirty limit. Thus, we will assume $\xi_e^{\rm Re} = \xi_0^{\rm Re}$.

Using the BS flux-flow model with the parameter estimates above results in a viscosity ratio $\eta_e^{\rm Re}/\eta_e^{\rm Al}=\left( \rho_n^{\rm Al}/\rho_n^{\rm Re} \right) \left( \xi_e^{\rm Al}/\xi_e^{\rm Re} \right)^2$ between $1$ and $4$, depending on the value for $\xi_e^{\rm Re}$, with the lower end of this range consistent with our measured viscosity ratio of $\sim 1$ from the $1/Q_v(B)$ data. 
Combining a viscosity ratio of $\eta_e^{\rm Re}/\eta_e^{\rm Al} \approx 1$ with the ratio of our depinning frequency fit values $f_d^{\rm Re}/f_d^{\rm Al}$ in Eq. (\ref{eq:fd-ratio}) results in $k_p^{\rm Re}/k_p^{\rm Al} \approx 5$.

The Re films in our experiment are nearly epitaxial, but highly twinned, based on RHEED measurements during the film deposition. Such extended defects likely result in strong pinning, particularly when the twins are oriented roughly along the length of the resonators, and thus perpendicular to the Lorentz force direction. 
On the other hand, the Al films deposited on non-heated substrates likely do not have such extended defect structures, but rather have defects that are small compared to $\xi_e^{\rm Al}$. Thus, one would expect weaker pinning in the Al films, consistent with $k_p^{\rm Re}/k_p^{\rm Al} > 1$. 
Because of the difference in pinning strength, flux creep is more significant in the Al films, particularly with $|B|$ well beyond $B_{th}$, where $\epsilon^{\rm Al} \approx 0.15$, compared to Re, where $\epsilon^{\rm Re} \approx 0.003$. 

\section{Threshold Fields and Vortex Distributions}

\begin{figure}
\centering
\includegraphics[width=3.35in]{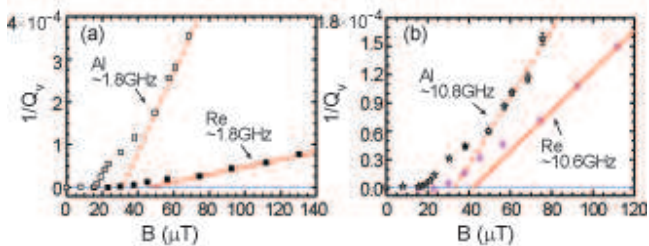}
\caption{(Color online) $1/Q_v(B)$ for $B \geq 0$ for Re and Al together for (a) lowest-frequency resonator; (b) highest-frequency resonator. Solid lines are linear fits to the $B$-dependence well beyond the shoulder region as described in text. $B_{th}$ corresponds to intercept of fit line with $1/Q_v=0$.
\label{fig:loss&shift-1.8-small-B}}
\end{figure}

In order to examine the field dependence near $B_{th}$ more closely,
in Figure \ref{fig:loss&shift-1.8-small-B} we plot $1/Q_v(B)$ for Re and Al together, but only for $B \geq 0$ for the lowest- [Fig. \ref{fig:loss&shift-1.8-small-B}(a)] and highest-frequency [Fig. \ref{fig:loss&shift-1.8-small-B}(b)] resonators. 
Near $B=0$ we observe $1/Q_v \approx 0$, indicating the presence of a threshold field below which vortices are not trapped in the resonators.
For magnetic fields 
beyond the initial onset from $1/Q_v=0$, there is a linear increase in $1/Q_v$ and  
we include linear fits to $1/Q_v(B)$ [Fig. \ref{fig:loss&shift-1.8-small-B}(a, b)].
Assuming $1/Q_v$ is proportional to the density of vortices in the resonator, our observed $1/Q_v(B)$ corresponds to a linear increase in vortex density with $B$, consistent with previous magnetic imaging measurements of field-cooled superconducting strips \cite{stan04,kuit08}.
Following this analysis, we can identify the point where these linear fits intercept $1/Q_v=0$ as $B_{th}$.
For the linear fits to the Re (Al) data, we obtain $B_{th}^{\rm Re}=45 \pm 2$~$\mu$T ($B_{th}^{\rm Al}=30 \pm 2$~$\mu$T) for the resonator near $1.8$~GHz.

The field-cooling of a thin superconducting strip has been studied theoretically by Likharev \cite{likharev72}, Clem \cite{clem98}, and Maksimova \cite{maksimova98} and these treatments were also described in Refs. \cite{stan04,kuit08,bronson06}. 
Sufficiently close to $T_c$, the effective thin-film penetration depth $\Lambda = 2 \lambda^2/d$ can become comparable to the strip width $w$, resulting in a uniform field distribution throughout the strip just below $T_c$. As the temperature is lowered further and superconducting order develops, the magnetic field through the strip nucleates into vortices and the ultimate spatial distribution of these depends on the vortex Gibbs free energy. 
The theoretical treatments of this problem have considered the Gibbs free energy for a single vortex in the strip, $G(x)$, where the $x$-coordinate is oriented across the width of the strip. This is determined by the interaction energy of the Meissner screening currents in the strip with the vortex and the self-energy of the vortex circulating currents. 
For small magnetic fields, $G(x)$ has a maximum in the center of the strip and falls off towards the edges of the strip, thus vortices do not nucleate in the strip upon cooling below $T_c$. 
As the strength of the cooling field is increased, the maximum in the middle of the strip flattens and eventually develops a dip in the center of the strip.  
Clem \cite{clem98} and Maksimova \cite{maksimova98} considered the development of this dip at $B_0 = \pi \Phi_0/4 w^2$ to correspond to the threshold field for trapping vortices near the center of the strip. 
Likharev argued that the trapping threshold is not reached until $G(x)=0$ in the center of the strip \cite{likharev72}, leading to the expression 
\begin{equation}
B_{s} = \frac{2 \Phi_0}{\pi w^2} \ln \left( \frac{\alpha w}{\xi} \right).
\label{eq:threshold}
\end{equation}
The constant $\alpha$ is related to the treatment of the vortex core and can be $2/\pi$ \cite{clem-unpub} or $1/4$ \cite{likharev72}. 
In their threshold field imaging measurements for Nb strips of different widths, Stan {\it et al.} found that Eq. (\ref{eq:threshold}) with $\alpha=2/\pi$ best described their observed values of $B_{th}$. 
A related model for vortex trapping in thin superconducting strips was proposed by Kuit {\it et al.} \cite{kuit08}, who considered the creation of vortex-antivortex pairs upon cooling through $T_c$. This model predicts a threshold field $B_K = 1.65 \Phi_0/w^2$ 
and successfully described the measured $B_{th}$ values for field-cooled YBCO strips of different widths \cite{kuit08}. 

If we take our $B_{th}$ values from the resonators near $1.8$~GHz, $B_{th}^{\rm Re}=45 \pm 2$~$\mu$T and $B_{th}^{\rm Al}=30 \pm 2$~$\mu$T, we can compare these with the various approaches. All of our resonators are nominally $12$~$\mu$m wide, thus $B_0=11$~$\mu$T and $B_K=24$~$\mu$T, which are below our measured $B_{th}$ and do not account for the differences between the Re and Al films. 
Applying Eq. (\ref{eq:threshold}) with $\alpha=2/\pi$ and assuming $B_s=B_{th}$, we obtain $\xi_e^{\rm Re}=60$~nm  and $\xi_e^{\rm Al}=360$~nm.  We note that these are within a factor of $\sim 2$ of our earlier estimates for $\xi_e^{\rm Re}$ and $\xi_e^{\rm Al}$, although one might expect the relevant coherence lengths in determining $B_{th}$ to be somewhat larger, corresponding to an elevated temperature at which vortices first become trapped in the films during the cooling process. 
On the other hand, the logarithm makes the dependence of $B_s$ on $w/\xi$ weak, thus making it difficult to perform a detailed quantitative comparison of $\xi_e$ based on the $B_{th}$ values alone. 
It is possible that the various trapping models require modifications to account for films of superconductors that are Type-I in the bulk and thus have relatively short penetration depths. If $\Lambda$ were to remain finite compared to $w$ at the temperatures where vortices begin to nucleate, the assumptions of weak screening and nearly uniform magnetic field distributions would need to be adjusted. Such a treatment is beyond the scope of our present work.

In Figure \ref{fig:loss&shift-1.8-small-B}(a, b) it is clear that $1/Q_v(B)$ deviates from the linear dependence for fields near $B_{th}$, and  $1/Q_v$ first becomes nonzero at $B=B_{onset}<B_{th}$. Thus $1/Q_v(B)$ exhibits shoulders, with $B_{onset}^{\rm Re} \approx 30$~$\mu$T and $B_{onset}^{\rm Al} \approx 20$~$\mu$T. 
Stan {\it et al.}  observed a deviation at small fields from the linear increase of vortex density with $B$ and the initial trapping of vortices occurred at magnetic fields somewhat below $B_{th}$ \cite{stan04}. 
This behavior was attributed to the presence of pinning, which resulted in local minima in $G(x)$ such that vortices could become trapped in the strip for $B<B_s$. Subsequently, Bronson {\it et al.} performed numerical simulations of this process and were able to obtain $n(B)$ curves that agreed with the measurements of Stan {\it et al.} \cite{bronson06}, where $n$ is the vortex density. By varying the pinning strength and density, Bronson {\it et al.} found regimes where $n(B)$ increased for $B<B_{th}$ with a smaller slope than the linear increase observed at large fields, as well as regimes where $n(B)$ increased more steeply for $B<B_{th}$ than the linear dependence at large fields, resulting in a shoulder on $n(B)$. 

If we assume that $1/Q_v(B)$ is proportional to $n(B)$ in our Re and Al resonators, the shoulders that we observe in $1/Q_v(B)$ for $B<B_{th}$ would be related to the same enhancement of vortex trapping by pinning as discussed in Refs. \cite{bronson06, stan04}.
However, if one examines the $r(B)$ data plotted in Fig. \ref{fig:r-param-field}, it appears that the situation may be somewhat more subtle. 
In the simplest case, if there were pinning wells of only one depth, one would expect $r(B)$ for a particular frequency to be flat, at least for $n(B)$ less than the density of pinning sites, as the $B-$dependence in the frequency shift and loss would cancel out for the calculation of $r$. On the other hand, a distribution of pinning well depths would likely favor the initial trapping of vortices in the deepest pinning wells, which would result in larger values of $r$ for $B$ just above $B_{onset}$. 
Such a picture, with a few deeper pinning wells, is consistent with our measurements of $r(B)$ in Re [Fig. \ref{fig:r-param-field}(a)], where $r(B)$ is mostly flat, with a small upturn as $|B|$ approaches $B_{onset}$, particularly for the lowest frequency resonator.
The measurements of $r(B)$ for Al [Fig. \ref{fig:r-param-field}(b)] exhibit a more gradual increase in $r$ as $|B|$ is reduced towards $B_{th}$, which may be related to a broader distribution of pinning energies in the Al films.
The decrease in $r(B)$ for $B_{onset}<B<B_{th}$ for the Al resonators, implying a weaker pinning of the initial vortices trapped in the film, 
is not understood presently.  

\begin{figure}
\centering
\includegraphics[width=3.35in]{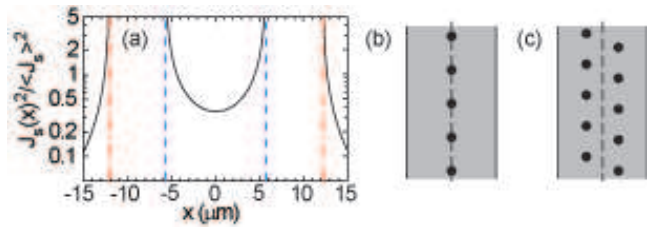}
\caption{(Color online) (a) Calculated current density distribution normalized by the average current density, $j(x)=J_s(x)^2/\langle J_s \rangle^2$, for CPW geometry with parameters for Al resonator: $w=11.5~\mu$m (indicated by blue dashed lines) and $6.4~\mu$m gap between the center conductor and the ground plane (indicated by red dash-dotted lines). Predicted vortex configurations in absence of pinning disorder based on Ref. \cite{bronson06} for (b) $B_{th} < B < 2.48 B_{th}$; (c) $B > 2.48 B_{th}$.
\label{fig:vortex-dist}}
\end{figure}

The vortex position in the resonator plays an important role in determining the response because of the non-uniform current density distribution in a superconducting coplanar waveguide $J_s(x)$, where the current density is larger at the edges. 
Thus, one must account for this when converting from $\tilde{\rho}_v$ to, for example, an effective resistance $R_v$, as in Eq. (\ref{eq:vortex-resistance}).
One approach for computing $J_s(x)$ involves numerically solving the two-dimensional London equations \cite{vanduzer81} for our CPW geometry and we plot this in Figure \ref{fig:vortex-dist}(a), where we have scaled $J_s(x)^2$ by the square of the average current density in the center conductor $\langle J_s \rangle^2$ to obtain the dimensionless factor $j(x)$ that we introduced previously. 
Vortices trapped along the centerline of the resonator will experience the smallest $J_s(x)$ and will thus 
exhibit the weakest response compared to vortices trapped near the resonator edge, which will respond most strongly.

From the vortex imaging measurements of Stan {\it et al.} and Kuit {\it et al.}, for $B$ just beyond $B_{onset}$, the vortices tended to line up in a single row along the centerline of the strips, while for somewhat larger $B$ the vortices formed multiple rows \cite{stan04, kuit08}. The numerical simulations of Bronson {\it et al.} indicated that the vortices should form a single row until $B=2.48 B_{th}$, at which point the distribution would split into two rows, one on either side of the strip centerline at $x \approx \pm (w/2)/3$ \cite{bronson06} [Fig. \ref{fig:vortex-dist}(b, c)]. For $B \approx 5 B_{th}$ the vortices would then form three rows, and so on. Following these simulations, our measured values of $B_{th}$ for the Re and Al films 
would correspond to the single-row configuration over much of the range of $B$ from our measurements, with the condition $B>2.48 B_{th}$ occurring towards the upper end of our cooling fields. 
Assuming a single-row configuration, we can estimate the typical vortex spacing near the middle of our field range if we assume the vortex density to be described by $n(B)=(B-B_{th})/\Phi_0$ for $B$ well beyond $B_{th}$, which is consistent with the measurements of Stan {\it et al.} \cite{stan04}. 
For a cooling field of $2 B_{th}$ as an example, this corresponds to a vortex spacing of $4$~$\mu$m at $B=86$~$\mu$T for the Re resonators. 

If pinning disorder were negligible, such that a clear transition from the single- to double-row configurations were to occur, one would expect a kink in the $1/Q_v(B)$ data with a larger slope at the largest fields of our measurements and beyond. 
The ratio of the slope of $1/Q_v(B)$ above and below the kink should correspond to the
ratio of $j(x)$ for $|x| =(w/2)/3 \approx 1.9~\mu$m (the vortex location in the two-row configuration) and $x=0$ (the vortex location in the one-row configuration), or $J_s (1.9$~$\mu$m$)^2/J_s (0)^2 = 1.15$. 
While such a kink is not clear from our data, a denser series of measurements over a somewhat larger field range could potentially reveal this slope change, provided the random pinning was not too strong.

Based on our measurements we have compiled a table of the various parameters for our Re and Al films (Table \ref{tb:Params}). 
These values can be used to compute $\tilde{\rho}_v$, then combined with $j(0)=0.35$ [Fig. \ref{fig:vortex-dist}(a)], assuming a single-row vortex configuration, and Eq.~(\ref{eq:f-over-loss-scaled}) to calculate $\partial \left( 1/Q_v \right)/\partial B$. This results in a calculated slope that ranges between a factor of 
$0.6 - 1.0$ 
of the fit slopes in Fig. \ref{fig:loss&shift-1.8-small-B} for the different Re and Al resonators. A similar analysis for the $\partial \left( \delta f/f_0 \right)/\partial B$ data, following the approach of Eqs. (\ref{eq:f-over-Q}-\ref{eq:f-over-loss-scaled}), yields a comparable level of agreement between our calculated and fit slopes. As described previously, disorder in the vortex positions caused by a random distribution of pinning sites could lead to deviations from the ideal single-row vortex configuration. Thus, this microwave vortex response model provides a satisfactory description of our measurements on Re and Al resonators. 
The same approach could be used to predict the microwave response of vortices in resonators patterned from other materials, although this would require some assumptions about the pinning strength in advance in order to estimate probable values for $f_d$ and $\epsilon$. \\

\begin{table}
\begin{tabular*}{0.46\textwidth}{@{\extracolsep{\fill}}|m{0.52in}|m{0.3in}|m{0.49in}|m{0.38in}|m{0.36in}|m{0.43in}|m{0.33in}|}
\hline
\hspace{0.1in}  
\multirow{2}{*} {Material} & $T_c$ & $\rho_n$ & $f_d$ & $\epsilon$ & $\xi_e$ & $B_{th}$ \\
& (K) & ($\mu \Omega$~cm) & (GHz) & & (nm) & ($\mu$T) \\
\hline \hline
Re & $1.70$ & $1.6$ & $22$ & $0.003$ & $50-100$ & $45$  \\
\hline
Al  & $1.13$ & $0.33$ & $4$ & $0.15$ & $230$ & $30$  \\
\hline
\end{tabular*}
\caption{Characteristic parameters for Re and Al thin films and vortices.}
\label{tb:Params}
\end{table}

\section{Conclusions}

We have measured the microwave response of vortices in superconducting thin films of Re and Al using resonant circuits. We introduced vortices by cooling in fields of the order of $100$~$\mu$T and below, and the vortex density exhibited a threshold field followed by a linear increase with field, consistent with previous vortex imaging experiments in Nb and YBCO strips. 
Despite the low vortex densities of our measurements, the response can be described reasonably in the context of an effective complex resistivity, which involves the pinning strength and vortex viscosity, along with a flux creep factor to account for the escape of vortices from pinning wells. 

Even at the small magnetic fields of our experiments, it is clear that the presence of vortices has a substantial influence on the resonator quality factor, although the film properties play an important role in the vortex response as well. 
Vortices in the Re resonators contribute significantly less loss, particularly at the lower frequencies of our measurements, compared to vortices in Al resonators. 
These differences are consistent with stronger pinning in the Re films relative to the Al. 
This suggests the possibility of controlling, and ideally reducing, the vortex loss in Al films with artificially patterned pinning configurations.
Nonetheless, it is important to design superconducting microwave circuits with narrow linewidths for large $B_{th}$ to eliminate trapped vortices due to ambient magnetic fields that are present when the devices are cooled through $T_c$. 
Of course, these ambient fields can be reduced in the first place with sufficient magnetic shielding. 
However, even with a low ambient field, one should also be careful to design a layout to minimize the possibility of pulsed control currents injecting vortices into resonators and other traces that require low loss.

\section{Acknowledgments}

We thank E. Silva for stimulating discussions. 
This work was supported by the National Science Foundation under Grant DMR-0547147. We acknowledge use of the Cornell NanoScale Facility, a member of the National Nanotechnology Infrastructure Network, which is supported by the National Science Foundation (Grant ECS-0335765), 

\bibliography{resonator}

\end{document}